# Beyond Nash Equilibrium in Open Spectrum Sharing: Lorenz Equilibrium in Discrete Games

Ligia C. Cremene
Department of Communications
Adaptive Systems Laboratory
Technical University of Cluj-Napoca, Romania
ligia.cremene@com.utcluj.ro

D. Dumitrescu
Department of Computer Science
Centre for the Study of Complexity
Babes-Bolyai University, Cluj-Napoca, Romania
ddumitr@cs.ubbcluj.ro

*Abstract*— **A new game theoretical solution concept for open spectrum sharing in cognitive radio (CR) environments is presented – the Lorenz equilibrium (LE). Both Nash and Pareto solution concepts have limitations when applied to real world problems. Nash equilibrium (NE) rarely ensures maximal payoff and it is frequently Pareto inefficient. The Pareto set is usually a large set of solutions, often too hard to process. The Lorenz equilibrium is a subset of Pareto efficient solutions that are equitable for all players and ensures a higher payoff than the Nash equilibrium. LE induces a selection criterion of NE, when several are present in a game (e.g. many-player discrete games) and when fairness is an issue. Besides being an effective NE selection criterion, the LE is an interesting game theoretical situation *per se*, useful for CR interaction analysis.**

## I. Introduction

The problem of finding an appealing solution concept for Open Spectrum Sharing (OSS) is addressed from a game theoretical perspective. *Lorenz equilibrium (LE)* [1] ensures a set of equitable Pareto optimal strategies [2] and a technique/criterion for selecting a Nash equilibrium (NE) when many are present.

OSS refers to spectrum sharing among secondary users (cognitive radios) in unlicensed spectrum bands [3]. Cognitive radio (CR) interactions are strategic interactions [4], [5]: the utility of one CR depends on the actions of all the other CRs in the environment. Game Theory (GT) provides a fertile framework and the tools for CR interaction analysis [4], [6], [7], [8], [9], [12], [13]. Insight may be gained on unanticipated situations that may arise in spectrum sharing, by devising GT simulations.

This paper tries to answer the question "Can we go beyond a Nash equilibrium?" [4] in one-shot spectrum sharing games. Usually, the outcome of a non-cooperative game is the Nash equilibrium - the most common solution concept. In OSS fairness is an issue [4], [12], [13] and usually, NE does not provide it.

Standard GT analysis of spectrum sharing is performed based on the continuous forms of the games. Yet, discrete modeling seems more realistic for spectrum access as users get to choose discrete quantities of the radio resources (number of channels, power levels, etc.). Discrete GT models usually yield multiple Nash equilibria. The number of NEa increases with the number of users. Equilibrium selection thus becomes an issue.

The basic assumptions of our approach are: *(i)* CRs are modelled as myopic, self-regarding players, *(ii)* CRs do not know in advance what actions the other CRs will choose, *(iii)* repeated interaction among the same CRs is not likely to occur on a regular basis [11], and *(iv)* CRs have perfect channel sensing and RF reconfiguration capabilities [9], [10]. Given these assumptions, one-shot, non-cooperative, discrete game analysis is considered relevant.

A well known game theoretical model – Cournot oligopoly [5] – is chosen as support for simulation, due to its simple and intuitive form and suitability for resource access modelling. Although it has been intensively used for spectrum trading modelling, the Cournot model has also been reformulated in terms of spectrum access [6], [12], [14] capturing general scenarios.

Considering the limitations of Nash and Pareto equilibria, a new solution concept is considered - the Lorenz equilibrium (a subset of Pareto optimal strategies). The Lorenz solution concept is a transformation that is applied to the payoffs (outcomes) of the game. It is suitable for both centralized and decentralized approaches to spectrum sharing. As Lorenz equilibrium is both equitable and Pareto efficient, it induces a NE selection criterion. Besides being an effective NE selection criterion, the LE is an interesting GT situation *per se*.

Numerical simulations reveal equilibrium situations that may be reached in simultaneous, open spectrum access scenarios. Lorenz equilibrium is detected and analyzed along with four other types of equilibria. Besides Nash and Pareto, new equilibrium situations establish, especially for *n*-player interactions. Heterogeneity of players is captured by joint Nash-Pareto equilibrium allowing CRs to exhibit different types of rationality [14].

The rest of the paper is structured as follows. Section II formally describes the Lorenz equilibrium. LE properties are discussed in Section III. Section IV introduces the LE generative relation as a basis for LE detection. Section V presents a general open spectrum access scenario for which the main equilibria are detected. Numerical simulations and results are discussed in Section VI. Section VII concludes our paper.

This paper was supported by CNCSIS –UEFISCDI, Romania, project TE 252/2010-2013.

## II. LORENZ EQUILIBRIUM

Generally, a game is defined as a system $G = (N, A_i, u_i, i = 1,...,n)$ [5] where:

(i) $N$ represents the set of $n$ players, $N = \{1,...,n\}$.

(ii) for each player $i \in N$, $A_i$ represents the set of actions $A_i = \{a_{i1}, a_{i2}, ..., a_{im}\}$; $A = A_1 \times A_2 \times ... A_n$ is the set of all possible game situations;

(iii) for each player $i \in N$, $u_i : A \to R$ represents the utility function (payoff).

A strategy profile is a vector $a = (a_1,...,a_n) \in A$, where $a_i \in A_i$ is an action of player $i$.

By $(a_i, a_{-i}^*)$ we denote the strategy profile obtained from $a^*$ by replacing the action of player $i$ with $a_i$, i.e. $(a_i, a_{-i}^*) = (a_1^*, a_2^*, ..., a_{i-1}^*, a_i, a_{i+1}^*, ..., a_1^*)$.

A strategy profile is said to be a Nash equilibrium if no player can improve her payoff by unilateral deviation [5].

Lorenz equilibrium is a new GT solution concept [1]. This solution concept is inspired by multicriteria optimization (MCO) where it is known as Lorenz dominance (LD) [15], or equitable dominance relation.

The standard solution concept in MCO is the Pareto set. A refinement of the Pareto dominance, LD is used in decision theory and fair optimization problems. The set of equitable efficient solutions is contained within the set of efficient solutions (Pareto). In GT, informally, the Pareto optimality (or Pareto efficiency) is a strategy profile so that no strategy can increase one player's payoff without decreasing any other player's payoff [2], [15].

In addition to the initial objective aiming at maximizing individual utilities, fairness refers to the idea of favoring well-balanced strategies [15]. Therefore, in fair optimization problems, we are interested in working with a preference relation $\succcurlyeq$ satisfying the following axioms:

(1) *P-Monotonicity*: For all $a', a'' \in A$:

$$u(a') \succcurlyeq_P u(a'') \Rightarrow u(a') \succcurlyeq u(a'') \quad (1)$$

and

$$u(a') \succ_P u(a'') \Rightarrow u(a') \succ u(a''), \quad (2)$$

where $\succcurlyeq_P$ ($\succ_P$) is the Pareto (Pareto strict) dominance relation [15].

(2) *Impartiality*: While dealing with uniform criteria, we want to focus on the distribution of outcome values while ignoring their ordering. In other words, a strategy generating individual payoffs: 4, 2, 0 for strategies $a_1, a_2, a_3$ respectively, should be considered equally as good as a strategy generating payoffs 0, 2, and 4. Hence it is assumed that the preference model is impartial (anonymous, symmetric). More formally:

$$u_{\tau(1)}(a), u_{\tau(2)}(a), ..., u_{\tau(n)}(a) \cong (u_1(a), u_2(a), ..., u_n(a)) \quad (3)$$

for any permutation $\tau$ of $\{1, 2, ..., n\}, a \in A$.

(3) *Principle of transfers*: The Pigou-Dalton principle of transfers [20] states that a transfer of any small amount from an outcome to any other relatively worse-off outcome results in a more preferred outcome vector. More formally:

$$u_i(a) > u_j(a)$$
$$\Rightarrow (u_1(a), ..., u_i(a) - \varepsilon, ..., u_j(a) + \varepsilon, ..., u_n(a)) \succ (u_1(a), ..., u_n(a))$$

for $0 < \varepsilon < u_{i'}(a) - u_{i''}(a)$. (4)

Thus a strategy generating all three individual payoffs equal to 2 is better than any strategy generating individual payoffs 4, 2, and 0.

The preference relation satisfying axioms (1) - (3) is called equitable (Lorenz) preference relation [15].

The relation of equitable dominance can be expressed as a vector inequality on the cumulative ordered payoffs. This can be mathematically formalized as follows. Let us consider the payoffs ordered in an ascending order

$$u_{(1)}(a) \leq u_{(2)}(a) \leq ... \leq u_{(n)}(a), a \in A \quad (5)$$

and define the quantities:

$$l_1(a) = u_{(1)}(a),$$
$$...$$
$$l_n(a) = \sum_{i=1}^{n} u_{(i)}(a). \quad (6)$$

Strategy $a$ is said to Lorenz dominate strategy $b$ (and we write $a \succ_L b$ if and only if [1], [15]:

$$l_i(a) \geq l_i(b), i = 1,...,n,$$
$$\exists j : l_j(a) > l_j(b). \quad (7)$$

*Lorenz equilibrium* of the game is the set of non-dominated strategies with respect to relation $\succ_L$ [1].

Informally, *Lorenz equilibrium is the set of the most balanced and equitable Pareto efficient strategies*.

## III. LORENZ EQUILIBRIUM PROPERTIES

In GT the Lorenz dominance relation is applied to address some limitations of standard GT solution concepts. Both Nash and Pareto equilibria have limitations when applied to real world problems. Nash equilibrium assumes players are rational agents choosing their strategies as best response to strategies chosen by other players. NE rarely ensures maximal payoffs for all players – it suffers from excessive competition among selfish players in a non-cooperative game, and the outcome may be inefficient [4]. On the other hand, Pareto equilibrium ensures the optimal payoffs but the set of Pareto-optimal strategies is often too large and too hard to process. Thus a decision making procedure is needed for selecting a particular Pareto optimal strategy. Moreover, payoffs of Pareto strategies may be highly unequal.

However, the LE provides a small subset of Pareto efficient strategies that are equitable for all players.

The advantage of the Lorenz equilibrium is that it preserves the qualities of Pareto equilibrium and the set of strategies is considerably smaller. Moreover the resulting strategies assure the maximal payoffs that are equitable for all players.

Lorenz equilibrium is characterized by equitable payoffs. In addition to the basic objective, aiming to maximize individual utilities, fairness refers to the idea of favouring well-balanced utility profiles.

Where multiple equilibria exist (in discrete games, for instance), the need for equilibrium selection arises. The main equilibrium selection criteria, as discussed in [4], are: Pareto optimality, equilibrium refinement, and evolutionary equilibrium.

If multiple Nash equilibria exist in a game (e.g. discrete many-player game), the closest ones to the Pareto optimal are usually chosen. But, usually, a large set (sometimes infinite) of Pareto optimal strategies exists. Considering the distance to the Lorenz set of optimal strategies, a more specific condition (criterion) is introduced. The selected NE will be closest to Lorenz equilibrium (Pareto optimal and also equitable strategy).

## IV. LORENZ EQUILIBRIUM DETECTION

An evolutionary method, based on generative relations [17], is used for equilibrium detection. An adaptation of the state-of-the-art Differential Evolution [19] is the underlying evolutionary technique. Other choices are also possible (e.g. an adaptation of NSGA-II [21]). The method is robust with respect to the nature of the game (continuous, discrete) and scalable to the number of players and the number of available resources (channels, power levels, etc.). It allows comparison of strategies and payoffs of several equilibria. The complexity is that of the underlying evolutionary techniques (e.g. [19], [21]).

In the framework of non-cooperative game theory, Lorenz equilibrium is defined using a generative relation based on the Lorenz dominance [1], [15].

*Generative relations* represent an algebraic tool for characterizing and detecting game equilibria [16], [17]. Generative relations are defined on the set of game strategies. The idea is that the non-dominated strategies with respect to the generative relation equals (or approximate) the equilibrium set.

Generative relations for Nash, Pareto, and joint Nash-Pareto equilibria may be defined [16]. The joint Nash-Pareto and Pareto-Nash equilibria applied to CR interaction analysis are discussed in [14], [18] for different game models.

*Generative relation of Lorenz equilibrium*. Let us consider a relation $R$ over $A \times A$. Strategy $(a', a'') \in R$ means strategy $a''$ dominates strategy $a'$.

A strategy $a'$ is non dominated with respect to relation $R$ if
$$\nexists a'' \in A: (a', a'') \in R$$

Let us denote by ND $R$ the set of non-dominated strategies with respect to relation $R$. A subset $A' \subset A$ is non-dominated with respect to $R$ if and only if $\forall a \in A', a \in NDR$.

Relation $R$ is said to be a generative relation for the equilibrium $E$ if and only if the set of non-dominated strategies with respect to $R$ equals the set $E$ of strategies (i.e. $NDR = E$).

Lorenz equilibrium of a game is the set of non-dominated strategies with respect to the $\succ_L$ relation as defined by (7). Therefore we may consider $\succ_L$ the generative relation for Lorenz equilibrium.

## V. OPEN SPECTRUM SHARING SCENARIO

In order to illustrate the detection and usefulness of the Lorenz equilibrium, a general open spectrum access scenario may be considered [6], [14]. The scenario is modelled as a non-cooperative, one-shot game. The discrete form of the game is analyzed, as it exhibits multiple Nash equilibria. Standard game models, in their continuous form ([4], [5], [8], [9], [12], [13]), do not capture the discrete nature of choices made by CRs. Therefore, we are lead to consider the discrete form games.

The players are $n$ CRs attempting to access a certain set of available channels (or whitespace) $W$. Each CR $i$ is free to implement a number of frequency hopping channels $a_i \in \{0, 1, ..., |W|\}$, where $a_i$ is a CR $i$ individual strategy and $|W|$ is the cardinality of the set $W$. A strategy profile is a vector $a = (a_1,...,a_n)$. Let us consider $K_i$ to be the rate of interfered symbols on each channel, and let $K_i \in [0, 10]$. This rate may also account for a range of factors that generally cause unused symbols (interference, noise, etc.) and it actually reflects the link-level performance on each channel.

Each CR is attempting to maximize its payoff $u_i$ given as:

$$u_i(a) = \left(|W| - \sum_{k=1}^{n} a_k\right) a_i - K_i a_i, i = 1, ..., n. \qquad (8)$$

The form of the chosen payoff function $u_i$ accounts for the difference between a function of goodput $\left(|W| - \sum_{k=1}^{n} a_k\right) a_i$ (a linear approximation of the number of non-interfered symbols per channel × number of frequency hopped channels) and the cost of simultaneously supporting $a_i$ channels: $K_i a_i$ (rate of interfered symbols per channel × number of channels). Other payoff functions may be considered (e.g. the ones in [12], [13]), as well as asymmetric costs.

The question for this general scenario is: how many simultaneous frequency hopping channels should each CR access in order to maximize its payoff in a stable game situation (equilibrium)?

Challenges of discrete games are related to: *(i)* computing Nash equilibria, *(ii)* existence of multiple NEa, and *(iii)* selection of an efficient NE (close to a Pareto optimal strategy or to Lorenz equilibrium).

## VI. EQUILIBRIUM DETECTION – NUMERICAL EXPERIMENTS

The effectiveness of the Lorenz equilibrium criterion for the Nash equilibrium selection problem becomes clear for many-player games (*n*-dimensional space). For the sake of accuracy and simplicity in illustrating the presence of LE, we have chosen to represent the two- and three-dimensional cases. Two and three CR simultaneous spectrum access scenarios are therefore considered. As the continuous modelling captures only partially the variety of possible equilibrium situations, challenging discrete instances of the game are considered.

Lorenz equilibrium is detected and analyzed along with four other types of equilibria: Nash, Pareto, and the joint Nash-Pareto, and Pareto-Nash equilibria.

Let us consider the following simulation parameters: $|W|=10$ (number of available channels) and $K_i = 1$ (equal unitary rate of interfered symbols per channel).

The reported results represent a sub-set of more extensive simulations. A population of 100 strategies has been evolved using a rank-based fitness assignment technique. In all experiments the process converges in less than 20 generations.

Fig. 1 illustrates the equilibrium strategies achieved by two CRs simultaneously trying to access the same set of ten available channels ($|W|=10$).

The computation of the NE strategy for the standard, continuous-form game – taken as a reference point – is straightforward and yields (3,3), the NE being unique [5]. NE is a stable strategy from which no CR has any incentive to individually deviate.

The discrete instance of the 2-player game reveals three Nash equilibria: (2,4), (3,3), (4,2) and one Lorenz equilibrium: (2,2) (Fig. 1 and Fig. 2). For a larger number of players even more NEa exist and LE is also a multi-element set.

The existence of multiple NEa indicates a certain degree of flexibility in choosing the number of channels to access – there are several situations from which the CRs have no incentive to unilaterally deviate.

The (3,3) NE strategy is the most stable game situation as it maintains even for the joint N-P and P-N strategies (they overlap, Fig.1). It is also the closest one to LE (2,2). As expected, the corresponding payoffs – NE: (9,9) – (Fig. 2), are the most equitable of the three NEa. The other two NEa, (2,4) and (4,2), are also stable and are maintained for one of the joint strategies (N-P or P-N), but are not equitable: payoffs are (6,12) and (12,6), respectively.

Fig. 2 illustrates the payoffs of the two players (CRs): $u_1(a_1, a_2)$ and $u_2(a_1, a_2)$. The three NE payoffs (6,12), (9,9), (12,6) offer a diversity of utilities among which one equitable solution: (9,9). LE payoff (10,10) is slightly higher than NE payoff (9,9). Yet, the corresponding number of accessed channels is smaller for LE: (2,2) than for NE: (3,3).

Fig. 3 and Fig. 4 capture the equilibrium situations (strategies and payoffs, respectively) for the discrete modelling of the 3-CRs simultaneous sharing of the same set of channels.

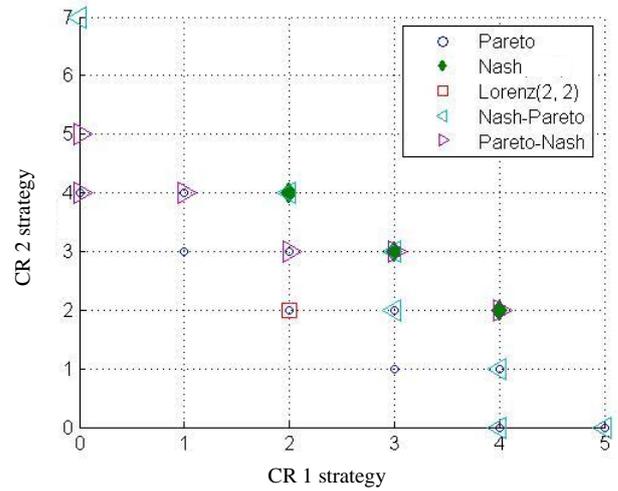

Figure 1. Discrete modelling. – two CRs (card W=10, K=1). Evolutionary detected equilibrium **strategies**: Nash: (2,4), (3,3), (4,2) Pareto, Nash-Pareto, Pareto-Nash, and Lorenz: (2,2).

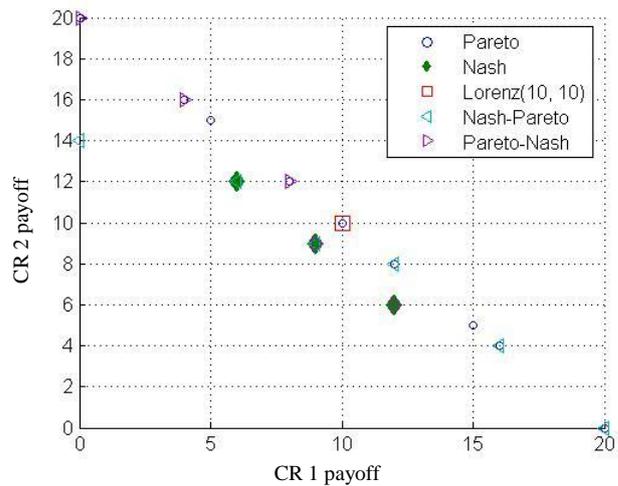

Figure 2. Discrete modelling – two CRs (card W=10, K=1). **Payoffs** of the evolutionary detected equilibria: Nash: (6,12), (9,9), (12,6), Pareto, N-P, P-N, and Lorenz: (10,10).

Seven Nash equilibria) are detected (Fig. 3): (2,2,2), (2,2,3), (2,3,2), (3,2,2), (1,3,3), (3,1,3), and (3,3,1). This indicates an even higher flexibility in choosing the number of accessed channels for each CR. Also the range of available payoffs is increased (Fig. 4). NE payoffs are (6, 6, 6), (4, 4, 6), (4, 6, 4), (6, 4, 4), (2, 6, 6), (6, 2, 6), (6, 6, 2). We may even notice one NE strategy (2,2,2) overlapping a LE and yielding identical payoffs (6,6,6). Obviously this is the preferred NE among the seven detected (according to the fairness criterion).

The effectiveness of LE criterion in NE selection is even more evident for *n*-player $(n \geq 3)$ games. In the *n-player* case the number of NEa increases polynomially. Moreover, the closeness to LE can no longer be indicated by visual inspection. The NE that is computationally detected as closest to LE may then be selected.

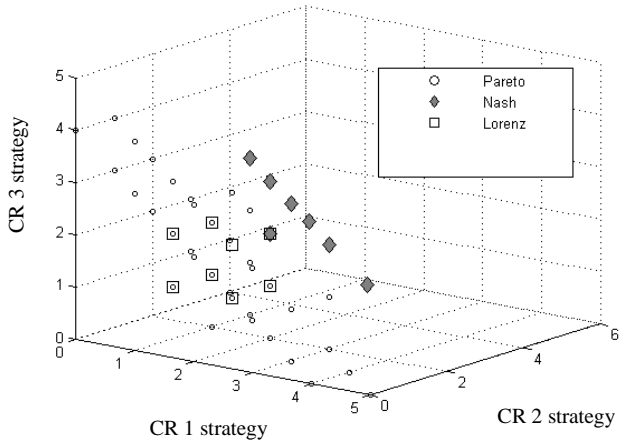

Figure 3. Three CR simultaneous access (card W=10, K=1). Discrete resource access modelling. **Strategies**: Nash: (2,2,2), (2,2,3), (2,3,2), (3,2,2), (1,3,3), (3,1,3), (3,3,1), Pareto, N-N-P, N-P-P, Lorenz: (2,2,2), (1,1,2), (1,2,1), (2,1,1), (2,2,1), (2,1,2), (1,2,2).

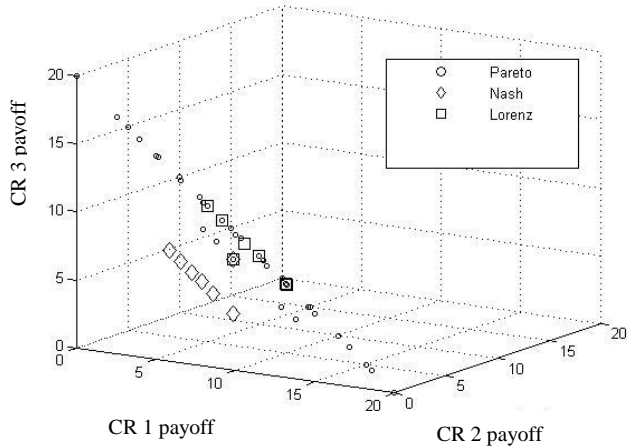

Figure 4. Three CR simultaneous access (card W=10, K=1). Discrete resource access modelling. **Payoffs**: Nash: (6, 6, 6), (4, 4, 6), (4, 6, 4), (6, 4, 4), (2, 6, 6), (6, 2, 6), (6, 6, 2), Pareto, N-N-P, N-P-P, and Lorenz: (6, 6, 6), (5, 5, 10), (5, 10, 5), (10, 5, 5), (8, 8, 4), (8, 4, 8), (4, 8, 8).

## VII. CONCLUSIONS

A new equilibrium concept for non-cooperative GT modelling of open spectrum sharing is considered: the Lorenz equilibrium. LE is a subset of Pareto optimal strategies that proves useful in selecting a NE when multiple ones exist (e.g. in many-player discrete games). LE is an appealing strategy concept for spectrum sharing games as it is both fair and profitable (usually ensures a higher payoff than NE and is Pareto optimal).